\documentclass[11pt]{article}
\usepackage{latexsym, amssymb, enumerate, amsmath,geometry}

\newcommand{\be}{\begin{equation}}
\newcommand{\ee}{\end{equation}}
\newcommand{\ben}{\begin{equation*}}
\newcommand{\een}{\end{equation*}}
\newcommand{\bea}{\begin{eqnarray}}
\newcommand{\eea}{\end{eqnarray}}
\newcommand{\beas}{\begin{eqnarray*}}
\newcommand{\eeas}{\end{eqnarray*}}

\def\R{\mathbb{R}}

\newtheorem{thm}{Theorem}[section]
\newtheorem{prop}[thm]{Proposition}
\newtheorem{lem}[thm]{Lemma}

\newtheorem{defn}[thm]{Definition}

\begin{document}

\title{Global classical solutions of the Vlasov-Darwin system for small initial data
}

\author{Martin Seehafer\thanks{University of Bayreuth, Department of Mathematics, D-95440 Bayreuth, Germany (martin.seehafer@uni-bayreuth.de).}
}

\maketitle

\begin{abstract}
A global-in-time existence theorem for classical solutions of the Vlasow-Darwin 
 system is given  under the assumption of smallness of the initial data. Furthermore it
is shown that in case of spherical symmetry the system degenerates to the relativistic Vlasov-Poisson system.

\end{abstract}

\section{Introduction}\label{intro}

\medskip
Kinetic models play an increasingly important role in todays plasma physics. 
On the one hand much effort is used to deepen our analytical understanding of some problems where no other description
seems to be adequate. On the other hand progress has also been achieved especially 
with numerical simulations (see e.g. \cite{sg06}).

In the kinetic picture, the particle distribution of a one-species plasma is described by a time dependent density function $f(t,x,p)$ on phase space.
If collisions of the particles are neglected and a relativistic model is used, then $f$ is subject to the transport equation
\be \label{vlasov1}
 \partial_tf +v(p)\nabla_xf + K(t,x)\nabla_p f = 0
\ee
with force term $K=E+v \times B$. Here $E$ and $B$ denote the electric and the magnetic field respectively
 and the relativistic velocity is given by
\be \label{rel_vel}
v(p) = \frac{1}{\sqrt{1+|p|^2}}p.
\ee
Note that all physical constants such as the speed of light or the rest mass of the particles have been set equal to unity.

Eq. (\ref{vlasov1}) is usually called the Vlasov equation. Expressions
for the charge and current densities $\rho$ and $j$ in terms of the phase space density $f$ are
given by
\be \label {rhoundj}
\rho(t,x)=\int f(t,x,p)dp, \quad j(t,x)= \int f(t,x,p)v(p)dp.
\ee
To obtain a self consistent closed system one has to take into account how the ensemble modeled by the density $f$ creates
the fields $E$ and $B$. Usually this is done with the full system of Maxwell's equations, but numerical difficulties of simulations
of that system have stimulated a search for alternatives (compare \cite{bms07}).
The present paper deals with what is known as the the Darwin approximation. Here the electric field is split
into a transverse and a longitudinal component as follows:
\be \label{ELundET}
E = E_L+E_T, \quad \nabla \times E_L = 0, \quad \nabla \cdot E_T = 0.
\ee
In the evolution part of the Maxwell equations the transverse part of the electric field is neglected, resulting in
\bea \label{evgl1}
\partial_t E_L-\nabla\times B = -j, \quad \nabla \cdot E_L = \rho \\
\label{evgl2}
\partial_t B+\nabla\times E_T = 0, \quad \nabla \cdot B = 0.
\eea
The system consisting of the equations (\ref{vlasov1}) -- 
 (\ref{evgl2}) is called the
\emph{Vlasov-Darwin} system. The main feature of this system is that the field equations are elliptic which in particular
facilitates a numerical treatment since a time integration step, which is needed to solve the Maxwell system can be avoided here (\cite{sg06}). The  justification of the model seems possible  in case the particle velocities are not too fast when compared
to the speed of light. 

Up to now there are only few mathematical results known for this system. In 2003 Benachour et al. \cite{ben03} proved an existence theorem for small initial data:
this assumption implies global-in-time existence of weak solutions of the Cauchy problem. Later Pallard \cite{pal06}
removed the smallness assumption and added a result about solvability of the Cauchy problem in a classical sense:
To a given initial datum $f_0\in C^2_c(\R^6)$ there exists $T>0$ and a classical solution $f\colon [0,T[\times \R^6 \rightarrow \R$ of the Vlasov-Darwin systems satisfying $f(0)=f_0$.

In the main part of the paper we present a result which is well known for the Vlasov-Poisson system (VP), the Relativistic Vlasov-Maxwell system (RVM), and other related systems such as the spherically symmetric Vlasov-Einstein system  (cf. \cite{bd85, gs87, rr92}) but seems to be new for the Vlasov-Darwin system: we consider
classical solutions of the Cauchy problem and show that these exist for all times if the initial data are chosen sufficiently small.
The precise statement of our result is contained in the next section, where we also formulate three propositions which are used to prove the theorem. 
Sections \ref{sources}--\ref{theorem} are devoted to proofs.
In the final Section \ref{symmetry} we take a look at spherically symmetric solutions. First it is shown that any symmetry of the initial datum  with respect to an orthogonal transformation is preserved for all times. This allows the conclusion that in case of spherical symmetry
the VD system reduces to the well known relativistic Vlasov-Poisson system. So in this case the solutions are global-in-time as well 
\cite{gs85}.

\section{Results\label{results}}

\medskip
Before presenting the main result of the present paper (which is formulated as the following theorem) we fix some notation.
Let $R_0,P_0>0$ be fixed throughout the paper.  For $r>0$ let $B_r:= B_r(0) = \{x \in \R^3 : |x| < r\}$.
Furthermore we specify the set where the initial data are taken from:
Let $C^2_c(\R^n)$ denote the space of twice continously differentiable functions on $\R^n$ with compact support and 
\ben
\mathcal{D}:= \{f \in C^2_c(\R^6) : f \geq 0, \|f\|_\infty \leq 1, \|\nabla f\|_{\infty} \leq 1, \text{supp } f \subset B_{R_0}(0)\times B_{P_0}(0) \}.
\een
The Legesgue space of square integrable functions is denoted by $L^2(\R^3)$ and $\mathbb{P}\colon L^2(\R^3) \rightarrow L^2(\R^3)$
is the projection on the divergence free part, which is discussed in the Appendix.

If $I$ is an intervall and $g \colon I \times \R^n \rightarrow \R^m$, we denote the quantity
$\sup_{x \in \R^n} |g(t,x)|$ by $\|g(t)\|_\infty$, and if  $K \subset \R^n$, the expression $\|g(t)\|_{\infty,K}$ means
 $\sup_{x \in K} |g(t,x)|$.

\medskip
\begin{thm}
There exists $\delta>0$ such that the classical solution of the VD system with initial datum $f_0$ in $\mathcal{D}$ satsifying $\|f_0\|_{\infty} \leq \delta$ exists globally in time.
\end{thm}

\medskip
For the proof of this result the reformulation of the field equations of the VD system in terms of potentials $\Phi$ and $A$ given in \cite{pal06} is used. Let 
\bea
\Delta \Phi = \rho, \quad \lim_{|x| \to \infty} \Phi(t,x)= 0, \\
\Delta A = - \mathbb{P}(j), \quad  \lim_{|x| \to\infty}A(t,x)=0. 
\eea
Then the components of the electromagnetic field are
\ben
E_L = \nabla \Phi, \quad B = \nabla \times A, \quad E_T = -\partial_t A,
\een
cf.  Lemma 2.3 in \cite{pal06}.

The proof of the theorem is given in Section \ref{theorem}. Sections \ref{sources}, \ref{fields}, \ref{dependence} contain the proofs of preliminary results, which are formulated
in Propositions \ref{prop1}, \ref{prop2} and \ref{prop3}. A prominent 
role in the following is played  by a certain decay condition. 

\medskip
\begin{defn} 
A classical solution $f \colon [0,T[\times \R^6 \rightarrow \R$  of the VD system is said to satisfy a free streaming condition with parameter $\alpha$ on an interval $[0,a]$ if 
\beas
\|E_T(t)\|_\infty + \|E_L(t)\|_\infty + \|B(t)\|_\infty \leq \alpha(1+t)^{-3/2}, \\
\|\nabla E_T(t)\|_\infty + \|\nabla E_L(t)\|_\infty + \|\nabla B(t)\|_\infty \leq \alpha(1+t)^{-5/2}
\eeas
for all $t \in [0,a]$.
\end{defn}

\medskip
As for the RVM there is a continuation criterion for solutions of the VD system, which says that solutions may be continued as long
as the momentum support $\{p \in \R^3 : \exists x,t\colon f(t,x,p) \neq 0\}$ remains bounded \cite{pal06}. Using this criterion one can show easily 
that a solution which satisfies a condition as the one above on its maximal interval of existence is indeed a global one.

\smallskip
We now discuss the main idea of the proof. The task will be to show that the free streaming condition implies decay 
of the source terms $\rho$ and $j$ as well as decay of the fields $E_L, E_T, B$. This will be done in the following
two propositions.

\medskip
\begin{prop} \label{prop1}
There exist $\alpha$,  $C_1(R_0,P_0)>0$ such that for every solution of the VD system with $f(0) \in \mathcal{D}$
which satisfies a free streaming condition on an interval  $[0,a]$ with parameter $\alpha$  it holds
\ben
\|\rho(t)\|_\infty  + \|j(t)\|_\infty\leq C_1|t|^{-3}, \quad \|\partial_x\rho(t)\|_{\infty} + \|\partial_xj(t)\|_{\infty}  \leq C_1.
\een
\end{prop}

\medskip
\begin{prop} \label{prop2}
There exist $\alpha$,  $C_2(R_0,P_0)>0$ such that for every solution of the VD system with $f(0) \in \mathcal{D}$
which satisfies a free streaming condition on an interval  $[0,a]$ with parameter $\alpha$  it holds for $t \in [1,a]$ 
\beas
\|E_T(t)\|_\infty + \|E_L(t)\|_\infty + \|B(t)\|_\infty \leq C_2 t^{-9/5}, \\
\|\nabla E_T(t)\|_\infty + \|\nabla E_L(t)\|_\infty + \|\nabla B(t)\|_\infty \leq C_2 t^{-8/3}
\eeas
\end{prop}

\medskip
These estimates provide the main ingredient needed for the bootstrap argument in the proof of the theorem.
It follows that a solution satisfying a free streaming condition decays asymptotically even faster. This is an
important point for the global existence argument.
To start
this bootstrapping
we need a further tool which is given in the next proposition: if the initial datum is chosen sufficiently small, then the fields
remain small for some time. This may be interpreted  as a statement about continous dependence
on initial data.

\medskip
\begin{prop} \label{prop3}
 Let $\epsilon, T >0$ be given. Then there exists $\delta >0$ such that any classical solution $f$ of the VD systems with 
$f(0) \in \mathcal{D}$ and $\|f(0)\|_\infty \leq \delta$ exists at least up to time $T$ and is such that
\ben
\|E_L(t)\|_\infty +\|E_T(t)\|_\infty + \|B(t)\|_\infty + \|\nabla E_L(t)\|_\infty+ \|\nabla E_T(t)\|_\infty+\|\nabla B(t)\|_\infty \leq \epsilon.
\een
\end{prop}

\newpage
\section{Decay of the source terms \label{sources}}

\medskip
For proving Proposition \ref{prop1}, we need the following

\medskip
\begin{lem} \label{gronwalltyplemma}
 Let $t>0$, $\xi \in C^2([0,t])$,  $\xi(t)=\dot{\xi}(t)=0$ and let
\ben
   |\ddot{\xi}(s)| \leq c_1(s)+c_2(s)|\xi(s)|+c_3(s)|\dot{\xi}(s)|
\een
for $s \in [0,t]$, where
 $c_1,c_2,c_3 \geq 0$ are continous and $c_3$ is monotonically decreasing.
Then 
\ben
 \left|\xi(s)\right| \leq \left(\int_s^t\sigma c_1(\sigma)d\sigma\right)
e^{\int_s^t\left(\sigma c_2(\sigma)+c_3(\sigma)\right)d\sigma}.
\een
\end{lem}

\medskip
{\em Proof. } Define $z(s):= \int_s^t|\dot{\xi}(\tau)|d\tau$, so that
$|\xi(s)| \leq z(s)$, $\dot{z}(s)=-|\dot{\xi}(s)|, z(t)=\dot{z}(t)=0$. Obviously
\begin{eqnarray*}
z(s) &=&    \int_s^t\left|\int_{\tau}^t\ddot{\xi}(\sigma) d\sigma\right|d\tau \\
    &\leq&  \int_s^t \int_{\tau}^t |\ddot{\xi}(\sigma)|d\sigma d\tau,
\end{eqnarray*}
so that by our assumptions
\begin{eqnarray*}
z(s) &\leq&   \int_s^t \int_{\tau}^t c_1(\sigma)  d\sigma d\tau + 
\int_s^t \int_{\tau}^t c_2(\sigma) z(\sigma)  d\sigma d\tau  -
\int_s^t \int_{\tau}^t c_3(\sigma)\dot{z}(\sigma)  d\sigma d\tau \\
& =:& I_1+I_2+I_3.
\end{eqnarray*} 
Changing the order of integration it follows that

\ben I_1 = \int_s^t\int_s^{\sigma} c_1(\sigma)d\tau d\sigma \leq \int_s^t \sigma c_1(\sigma) d\sigma
\een as well as
\ben I_2 = \int_s^t \int_s^{\sigma} c_2(\sigma) z(\sigma)  d\tau d\sigma  \leq 
\int_s^t \sigma c_2(\sigma) z(\sigma)  d\sigma.
\een
The integral $I_3$ can be estimated in the following way:
\ben I_3 \leq \int_s^t c_3(\tau)\left(-\int_{\tau}^t \dot{z}(\sigma)  d\sigma\right) d\tau
= \int_s^t c_3(\tau) z(\tau)  d\tau,
\een
where the monotonicity of $c_3$ was used in the first step and the relation $z(t)=0$ in the second.
Hence the function $z$ satisfies the integral inequality
\ben
z(s) \leq \int_s^t \sigma c_1(\sigma) d\sigma + \int_s^t \left[\sigma c_2(\sigma) + c_3(\sigma)\right]z(\sigma) d\sigma,
\een
so that by Gronwall's Lemma
\ben
|\xi(s)| \leq z(s) \leq \left(\int_s^t \sigma c_1(\sigma) d\sigma  \right) 
e^{\int_s^t \left[\sigma c_2(\sigma) + c_3(\sigma)\right] d\sigma}.
\een
\hfill $\Box$

\bigskip
{\em Proof of Proposition \ref{prop1}.} 

The proof presented here is an adaptation of the corresponding argument for the Vlasov-Poisson system (cf. \cite{r07}).
To get decay of $\rho$ a change of variables is performed in the integral defining it. The transformation determinant
appearing can be shown to decay fast enough.

Let $f$ be a classical solution of the VD system with $f(0) = f^\circ \in \mathcal{D}$ and denote by $(X(s,t,x,p)$, $P(s,t,x,p))$
the corresponding solution of the characteristic system
\beas
	\dot X(s,t,x,p)&=&v(P(s,t,x,p)), \\
	\dot P(s,t,x,p)&=&E(X(s,t,x,p),s)+v(P(s,t,x,p)) \times B(X(s,t,x,p),s)
\eeas
with initial condition $X(t,t,x,p)=x, P(t,t,x,p)=p$. 

Then we have
\ben
f(t,x,p)=f^\circ(X(0,t,x,p),P(0,t,x,p)).
\een
If  $|p| \leq P_0$, then by the free streaming condition
\ben
|P(t,0,x,p)| \leq P_0 + \int_0^t\left( \|E(s)\|_\infty + \|B(s)\|_\infty\right) ds \leq P_0 +\alpha\int_0^t(1+s)^{-3/2}ds \leq P_0+2\alpha.
\een
Hence for small enough $\alpha$ we may conclude that $|p|\geq P_0+1$ implies $f(t,x,p)=0$.

Define
\ben \xi (s):= \partial_pX(s,t,x,p)-(s-t)Dv(p).
\een
We have $\xi(t)=0$, and using the characteristic system one obtains
\ben
\dot{\xi}(s) = Dv(P(s))\partial_pP(s) - Dv(p).
\een
So $\dot\xi (t)=0$. Differentiating once more we get
\ben
\ddot{\xi}(s) = D^2v(P(s))\dot{P}(s)\partial_pP(s)+Dv(P(s)) \partial_p\dot{P}(s).
\een
As is easily checked, $Dv(p)$ and $D^2v(p)$ are bounded independently of $p$, so that
\ben
|\ddot{\xi}(s)| \leq C\left(|\dot P(s)| |\partial_pP(s)| + |\partial_p\dot P(s)| \right)
\een
and therefore by the characteristic system
\ben
|\ddot{\xi}(s)| \leq
C\left(
|\partial_pP(s)|(\|E(s)\|_\infty + \|B(s)\|_\infty) + (\|\nabla E(s)\|_\infty + \|\nabla B(s)\|_\infty) |\partial_pX(s)|
\right).
\een
Resubstituting we have 
$$|\partial_pX(s)| \leq |\xi(s) + (s-t)Dv(p)|$$
 and
$$\partial_pP(s) = \left(Dv(P(s))\right)^{-1}(\dot \xi(s) + D_pv).$$
Assuming $|p| \leq P_0+1$, we can estimate
\ben
|\partial_pP(s)| \leq C(P_0)\left(|\dot \xi(s)|+1\right).
\een
Using the free streaming condition, we finally obtain the following second order differential inequality for $\xi$:
\ben
|\ddot{\xi}(s)| \leq C(P_0)\alpha
\left\{(1+s)^{-3/2} + (t-s)(1+s)^{-5/2} + (1+s)^{-3/2}|\dot \xi(s)| +(1+s)^{-5/2}|\xi(s)|
\right\}
\een
By Lemma \ref{gronwalltyplemma} 
\ben
|\xi(s)| \leq C(P_0)\alpha(t-s)e^{C(P_0)\alpha},
\een
where we possibly have to adjust the constant $C(P_0)$. In terms
the characteristic variables this means for $\alpha$ chosen sufficiently small that
\ben
|\partial_pX(0,t,x,p) + tDv(p)| \leq \epsilon t,
\een
where $\epsilon>0$ is prescribed such that 
\ben
\frac{|p\otimes p|}{1+|p|^2}+\epsilon \sqrt{1+|p|^2} \leq \beta <1 \text{ for } |p| \leq P_0+1.
\een
Here $p\otimes p$ denotes the matrix whose $(i,j)$-entry is $p_ip_j$. This implies
\ben
\left|\partial_pX(0,t,x,p) + \frac{t}{\sqrt{1+|p|^2}}I\right| \leq \epsilon t + \left|\frac{t}{\sqrt{1+|p|^2}}I-tDv(p)\right|.
\een
By direct computation 
$$Dv(p)=\frac{1}{\sqrt{1+|p|^2}}\left(I-\frac{p\otimes p}{1+|p|^2}\right),$$
hence
$$\left|\partial_pX(0,t,x,p) + \frac{t}{\sqrt{1+|p|^2}}I\right| 
\leq \epsilon t + \frac{t}{\sqrt{1+|p|^2}}\frac{|p\otimes p|}{1+|p|^2}< \frac{t}{\sqrt{1+|p|^2}}\beta.
$$
So the linear map $\partial_pX(0,t,x,p)$ is invertible and in conclusion the transformation $\Psi\colon B_{P_0+1} \rightarrow \R^3, v \mapsto X(0,t,x,p)$ is a local diffeomorphism.
It is even a diffeomorphism onto its image, since it is ono-to-one as well: Let $p, \bar p \in B_{P_0+1}$ be given and 
$p_\tau:= \tau p + (1-\tau)\bar p$. Then
\beas
|\Psi(p)-\Psi(\bar p)| &=& \left| \int_0^1 \partial_p X(0,t,x,p_\tau)(p-\bar p)d\tau\right| \\
&= & \left| \int_0^1 \left[\partial_p X(0,t,x,p_\tau)+\frac{t}{\sqrt{1+|p_{\tau}|^2}} I \right](p-\bar p)d\tau -  \int_0^1\frac{t}{\sqrt{1+|p_{\tau}|^2}} (p-\bar p)d\tau\right| \\
& \geq & t|p-\bar p|\int_0^1\frac{1}{\sqrt{1+|p_\tau|^2}}d\tau -\beta t |p-\bar p| \int_0^1\frac{1}{\sqrt{1+|p_\tau|^2}} d\tau \\
&\geq & C(1-\beta)t|p-\bar p |,
\eeas
where $C$ depends only on $P_0$.

Denote the open range of $\Psi$ by $U$ and let $\Phi\colon U \rightarrow B_{P_0+1}$ be its inverse.
Calculation gives
\beas
 \rho(t,x) &=& \int_{B_{P_0+1}}f(t,x,p)dp \\
& =& \int_{\Phi(\Psi(B_{P_0+1}))}f^\circ(\Psi(p),P(0,t,x,p))dp \\
&=& \int_{\Psi(B_{P_0+1})} f^\circ(w,P(0,t,x,\Phi(w))) |\det D\Phi(w)| dw.
\eeas
For the functional determinant showing up here we have by our previous calculations
\be \label{fundet}
|\det D\Phi(w)| =|\det [D\Psi(\Phi(w))]^{-1}|= \frac{1}{|\det D\Psi(\Phi(w))|} .
\ee
Note that 
\beas
\det D\Psi (p)  &=& \det\left(\partial_pX(0,t,x,p)+\frac{t}{\sqrt{1+|p|^2}}I-\frac{t}{\sqrt{1+|p|^2}}I\right)  \\
& = & \frac{t^3}{(1+|p|^2)^{3/2}}\det \left(\sqrt{1+|p|^2} \frac{\partial_pX(0,t,x,p) + \frac{t}{\sqrt{1+|p|^2}}I}{t} - I \right).
\eeas
But for the first matrix in the argument of the determinant we have 
$$\left| \sqrt{1+|p|^2} \frac{\partial_pX(0,t,x,p) + \frac{t}{\sqrt{1+|p|^2}}I}{t}\right| \leq \beta <1,$$
so the absolute value of the determinant is bounded from below by a positive constant $C_\beta$.
Returning to (\ref{fundet}), it is seen that
$$ |\det D\Phi(w)| \leq \frac{C_{\beta,P_0}}{t^3},$$
resulting in
\ben \rho(t,x) \leq \frac{CR_0^3\|f^\circ\|_{\infty}}{t^3}.
\een
In addition
\ben
|j(t,x)|  \leq \rho(t,x) \leq  \frac{C^*}{t^3}.
\een
For the bounds to be obtained for $\partial_x \rho$ and $\partial_x j$ note that
\beas
|\partial_x\rho(t,x)| &\leq& C(P_0+1)^3\|\partial_xf(t)\|_{\infty}, \\ 
|\partial_xf(t,x,p)|&\leq& C(|\partial_xX(0,t,x,p)| + |\partial_xP(0,t,x,p)|).
\eeas
Next let  $\xi(s):= \partial_xX(s,t,x,v) -\text{id}$ such that
$\dot{\xi}(s) = Dv(P(s)) \partial_xP(s)$. Obviously $\xi(t)=\dot \xi (t)=0$.
Differentiating further one has
\beas
|\ddot{\xi}(s)| 
&=& |D^2v(P(s)) \dot P(s,t,x,p)\partial_xP(s,t,x,p) + Dv(P(s)) \partial_x \dot P(s,t,x,p)| \\
&\leq& \alpha C\left( (1+s)^{-3/2}|\partial_xP(s,t,x,p)| + (1+s)^{-5/2}|\partial_x X(s,t,x,p)|\right),
\eeas
where again the decay of the fields due to (almost) free streaming was employed.
By definition
\ben
|\partial_xX(s) \leq |\xi(s)|+1 \quad \text{and}\quad |\partial_x P(s)| \leq C|\dot \xi(s)|
\een
and we may assume $|p| \leq P_0+1$ to discover the relation
\ben
|\ddot{\xi}(s)|\leq C\alpha \left\{ 
(1+s)^{-5/2} |\xi(s)| + 
(1+s)^{-3/2} |\dot{\xi}(s)| + \alpha(1+s)^{-5/2}
\right\},
\een
which by Lemma \ref{gronwalltyplemma} implies
\ben
|\xi(s)| \leq 
C\alpha\int_s^t(1+\sigma)^{-3/2}d\sigma e^{C\alpha\int_s^t(1+\sigma)^{-3/2}d\sigma} 
\leq 2C\alpha e^{2C\alpha}.
\een
An easy application of Gronwall's Lemma shows that $|\dot \xi (s)|$ is bounded too, which means
\ben
(|\partial_xX(0,t,x,p)| + |\partial_xP(0,t,x,p)|) \leq C
\een
and all claims are proved.
\hfill $\Box$

\section{Decay of the fields \label{fields}}

\medskip
{\em Proof of Proposition \ref{prop2}.} First let the constants $\alpha$ and $C_1$ be as given by Proposition \ref{prop1}. We want to get sufficiently good decay rate estimates for the fields from the field equations, the decay of the source terms, and the free streaming condition. The field $E_T$ is treated first. We have 
$$\Delta E_T = -\partial_t(\Delta A) = \partial_t (\mathbb{P} j) = \mathbb{P}(\partial_t j),$$
where the last equation, i.e. the commutativity of $\mathbb{P}$ and $\partial_t$, is read off directly from the Fourier representation of the projection operator, compare (\ref{fourier_P}) or the Appendix.

The Vlasov equation, Eq. (\ref{vlasov1}), then implies
\ben
 \partial_tj(t,x)=\int \partial_tf(t,x,p)v(p)dp = -\int \left<v(p), \nabla_xf\right> v(p) dp - \int v(p) \otimes K(t,x,p) \nabla_pf dp.
\een
Here $K(t,x,p)=E(t,x)+v(p)\times B(t,x)$. Integration by parts in the last term finally leads to
\beas
\partial_t j(t,x) &=& - \int \mathrm{div}_x[f(t,x,p)v(p)\otimes v(p)]dp +\int \frac{I-v\otimes v}{\sqrt{1+p^2}}f(t,x,p)K(t,x,p)dp \\
                &=:& G_1(t,x)+G_2(t,x),
\eeas
where the divergence appearing is to be understood row-wise. 

Writing $E_T = E^1_T+E^2_T$, where the components of the r.h.s. are solutions of
$\Delta E^1_T = \mathbb{P}(G_1)$ and $\Delta E^2_T = \mathbb{P}(G_2)$ respectively,
we treat each of them seperately. Recall the Fourier representation of the projection operator $\mathbb{P}$:
\bea \label{fourier_P}
 \mathbb{P}F(x)&=&\int e^{ikx}\frac{|k|^2I-k\otimes k}{|k|^2}\hat{F}(k)dk, \\
\hat F(k)&=& (2\pi)^{-3}\int e^{-ikx}F(x)dx
\eea
(compare the Appendix). The solution of the Poisson equation may then be expressed as
\ben
 E^l_T(t,x)=-\int e^{ikx}\frac{|k|^2I-k\otimes k}{|k|^4}\hat{G}_l(k)dk, \quad l=1,2.
\een
Introducing $M = (M_1,M_2,M_3)= \int f(t,x,p)v(p)\otimes v(p) dp$, it comes
\beas
E^1_T(t,x) &=& \int e^{ikx}\frac{|k^2|I-k\otimes k}{|k|^4}\hat{G}_1(t,k)dk  \\
&& = \sum_j \int e^{ikx}\frac{|k^2|I-k\otimes k}{|k|^4}ik_j\hat{M}_j(t,k)dk  \\
&& = i \sum_j\int e^{ikx}m_j(k)\hat{M}_j(t,k)dk,
\eeas
where $m_j(k)=\frac{|k^2|I-k\otimes k}{|k|^3}\frac{k_j}{|k|}$ is a function homogeneous of degree -1.
\/ The theory of pseudo differential operators (compare (45) and the proof of (47) in \cite{ks02}) permits us to estimate as follows:
\bea \label{ET1bound}
\|E^1_T(t)\|_\infty & \leq & C(\|M\|_{\infty} + \|M\|_p), \quad  1 \leq p <3, \\
\label{nablaET1bound}
\|\nabla_x E^1_T(t)\|_\infty & \leq &
C(\gamma^{3/p'-2}\|\nabla M\|_p + \log(\gamma^{-1})\|M\|_{\infty} +\|M\|_q),
\eea
 $3<p < \infty, 1<q< \infty$.
Here the parameter $\gamma$ is restricted to the interval $]0,1]$.
Since $\|M(t)\|_{\infty} \leq  \|\rho(t)\|_\infty \leq C t^{-3}$,
the first equation implies setting $p=5/2$:
\ben \|E^1_T(t)\|_\infty \leq Ct^{-9/5}\een
Using estimates from the proof of Proposition \ref{prop1} and $\|\nabla M(t)\|_{\infty} \leq \|\nabla \rho(t)\|_\infty \leq C$ 
it  follows that
\ben
 \|\nabla M(t)\|_p \leq \|\nabla M\|^{\frac{p-1}{p}}_\infty\|\nabla M\|^{1/p}_1 \leq C\|\nabla M\|^{1/p}_1 \leq Ct^{3/p},
\een
so that setting $\gamma = t^{-3}$ we get for $t\geq 1$:
\ben
 \|\nabla E^1_T(t)\|_\infty \leq 
C\left(
t^{-3(1-3/p)+3/p}+t^{-3}\log t+t^{-3(q-1)}
\right)
\leq C t^{-8/3}.
\een
(here the choice $p=36, q = 17/9$ was made).

Now consider $\Delta E^2_T = \mathbb{P}(G_2)$. We have
\beas
 E^2_T(t,x) = \int e^{ikx}\frac{|k|^2I-k\otimes k}{|k|^4}\hat{G}_2(t,k)dk =  \int e^{ikx}m_{-2}\hat{G}_2(t,k)dk, \\
 \partial_jE^2_T(t,x) = i\int e^{ikx}\frac{|k|^2I-k\otimes k}{|k|^3}\frac{k_j}{|k|}\hat{G}_2(t,k)dk = \int e^{ikx}m_{-1}\hat{G}_2(t,k)dk .
\eeas
The symbols $m_\alpha$ showing up here are homogeneous of degree $\alpha$.
\/ A simple adaptation of the proof of (45) in \cite{ks02} shows that
\be \label{ET2bound}
 \|E^2_T(t)\|_{\infty} \leq C( \|G_2(t)\|_\infty +\|G_2(t)\|_p ), \quad 1 \leq p < 3/2,
\ee
and as before
\be \label{nablaET2bound}
\|\partial_x E^2_T(t)\|_{\infty} \leq C( \|G_2(t)\|_\infty +\|G_2(t)\|_p ), \quad 1 \leq p < 3.
\ee
Since $|G_2(t,x)| \leq \|K(t)\|_{\infty}\rho(t,x)$, we have
\ben
\|G_2(t)\|_{5/4} 
\leq 
C(1+t)^{-3/2}\|\rho(t)\|^{1/5}_\infty \|\rho(t)\|^{4/5}_1 \leq t^{-21/10}
\een
and
\ben
\|G_2(t)\|_{5/2} \leq
C(1+t)^{-3/2}\|\rho(t)\|^{3/5}_\infty \|\rho(t)\|^{2/5}_1 \leq
 t^{-33/10}.
\een
Altogether this implies
\beas
\|E_T(t)\|_{\infty} &\leq & Ct^{-9/5}, \\
\|\nabla_x E_T(t)\|_{\infty} & \leq & Ct^{-8/3},  \qquad t \geq 1.
\eeas
Now we come to the other fields. The longitudinal part $E_L$ of the electric field is treated exactly as in the case of the Vlasov-Poisson system (cf. \cite{r07}):
\beas
\|E_L(t)\|_{\infty} & \leq & Ct^{-2},\\
\|\partial_x E_L(t)\|_{\infty}  & \leq & Ct^{-3}\log t.
\eeas

The bounds for the magnetic field $B = \nabla  \times A$ field are obtained in a way analogous to the procedure used so far.
First we have a representation
\ben
\nabla A(t,x) = \int e^{ikx}m_{-1}(k)\hat{j}(t,k)dk
\een
with $m_{-1}$ homogeneous of degree $-1$. Therefore
\ben
\|B(t)\|_\infty \leq \|j(t)\|_\infty +\|j(t)\|_{5/2} \leq Ct^{-9/5}.
\een
In analogy to our treatment of  $E_T$ we find
\be \label{nablaBbound}
\| \nabla_x B(t) \|_\infty \leq C
\left(\gamma^{3/p'-2}\|\nabla j\|_p + \log(\gamma^{-1})\|j\|_{\infty} +\|j\|_q
 \right)
\ee
and the proof may be completed as shown before. \hfill $\Box$

\section{Continous dependence\label{dependence}}

\medskip
In this section we denote by $C$ a constant depending only on $R_0,P_0$, which may change from
line to line. For the proof we collect some facts first.


Let $(f, E_L, E_T, B)$ be a solution of the VD system on some time interval $[0,T[$ with $f_0 = f(0) \in \mathcal{D}$. 
Define
\ben
Q(t):= \sup \{|p| : \exists x, 0\leq s\leq t\colon f(t,x,p) \neq 0\}.
\een

Then we have the following

\medskip
\begin{lem} Let $f$ be a solution with $f(0) \in \mathcal{D}$. Then there holds
\beas
\| \rho(t) \|_{4/3} + \|j(t)\|_{4/3} &\leq& C\|f_0\|_\infty, \\
\|A(t)\|_\infty & \leq & C \|f_0\|Q(t)^{1/3}, \\
\|\nabla A(t)\|_\infty + \|\nabla \Phi(t)\|_\infty & \leq & C\|f_0\|_\infty Q(t)^{5/3}.
\eeas
\end{lem}

\medskip
To estimate the field $E_T$ only a local result is available.

\medskip
\begin{lem} \label{localboundET}
Let $f \colon [0,T]\times \R^6 \rightarrow \R$ be a solution with $f(0) \in \mathcal{D}$. 
Then
\ben
\|E_T(t)\|_{\infty, B_{R_0+T}} \leq C_{R_0+T}(1+\|\rho(t)\|_3)(\|F(t)\|_{6/5} + \|F(t)\|_2),
\een
where $F = F_1 +F_2$ with
\beas
F_1(t,x) &=& \int \mathrm{div}_{(x)} (f(t,x,p)v(p)\otimes v(p))dp,\\
F_2(t,x) &=& \int \frac{I-v(p)\otimes v(p)}{\sqrt{1+|p|^2}}f(t,x,p)(E_L(t,x)+v(p)\times B(t,x))dp.
\eeas
\end{lem}

\medskip
Detailed proofs of these Lemmas are 
given in Pallard's paper \cite{pal06}, where also the following theorem is proved. 

\medskip
\begin{thm} Let $f_0 \in C^2(\R^6)$. Then there exists $T^*>0$ and a unique solution $(f,E_L,E_T,B)$ 
to the Vlasow-Darwin system with $f(0) = f_0$  satisfying
\beas
 f & \in & C^1([0,T^*[ \times \R^3 \times \R^3), \\
E_L,B & \in & C^1([0,T^*[ \times \R^3), \\
E_T,\nabla_xE_T & \in & C([0,T^*[ \times \R^3),
\eeas
and such that for any $t \in [0,T^*[$ the distribution function $f(t,.)$ is compactly supported.
\end{thm}

\medskip
By inspection of the constants in the proof one finds that a strict lower bound for
$T^*$ is given by $T':=(C\|f_0\|^2_\infty)^{-1}$ with a constant $C$ independent of $f_0$, i.e $T^*>T'$. In addition one has
\be \label{localboundQ}
Q(t) \leq C(R_0,P_0) \qquad (0 \leq t \leq T').
\ee
One last ingredient for the proof of Proposition \ref{prop3} is contained in the following

\medskip
\begin{lem} Let $f$ be a solution and $T'$ be defined as above. Then
\ben
\|\nabla_{x,p}f(t)\|_\infty \leq C(R_0,P_0) \qquad (0 \leq t \leq T').
\een
\end{lem}

\medskip
The proof can again be found in \cite{pal06}. From the Lemma we immediately deduce the
bounds
\ben \|\nabla \rho(t)\|_\infty + \|\nabla j(t)\|_\infty  \leq C, \qquad 0\leq t \leq T'.\een
 With these tools at hand we are now ready for the

\bigskip
{\em Proof of Proposition \ref{prop3}.}

Let $\epsilon, T>0$ be given. From the above facts one finds immediately that the solution interval
can be made as long as we wish and that 
$\|E_L(t)\|_\infty$ and $\|B(t)\|_\infty$ can be made as small as necessary
by choosing $\delta$ sufficiently small.

It is standard to obtain a bound for $\|\nabla_xE_L(t)\|_\infty$ (see \cite{r07}), and for
 $\|\nabla B(t)\|_\infty$ we can use Eq. (\ref{nablaBbound}) from Section \ref{fields}:
By finite propagation speed and since $\|\nabla j(t)\|_\infty$ remains bounded for
$t \in [0,T']$, we can choose the parameter $\gamma$ on the right-hand side of the inequality properly to get the result. 

We still have to get control over $\|E_T\|_\infty$ and $\|\nabla E_T\|_\infty$. From (\ref{ET1bound}) and 
(\ref{ET2bound}) we have the estimate
\ben
\|E_T(t)\|_\infty \leq C(\|\rho(t)\|_\infty+ \|\rho(t)\|_2 +\|G_2(t)\|_\infty +\|G_2(t)\|_{5/4}),
\een
where the notation introduced in Section \ref{fields} is used again. Now
\beas
|G_2(t,x)| & \leq & \left| \int \frac{I-v(p)\otimes v(p)}{\sqrt{1+|p|^2}} f(t,x,p)K(t,x,p)dp \right| \\
& \leq & C \int_{|p|\leq Q(t)} f(t,x,p) |K(t,x,p)| dp \\
&\leq & CQ(t)^3 \|f_0\|_\infty \|K(t)\|_{\infty, B_{R_0+T}} \chi_{B_{R_0+T}}(x).
\eeas
and since we have bounds  for  $\|K(t)\|_{\infty, B_{R_0+T}}$ by  Lemma \ref{localboundET}, we get 
$\|E_T(t)\|_\infty  \leq \epsilon$ for $\|f_0\|_\infty$ chosen sufficiently small in view of (\ref{localboundQ}).

To estimate $\| \nabla E_T(t) \|$ we note as in Section \ref{fields}:
\ben
\|\nabla_x E_T(t)\|_\infty \leq 
C\left(\gamma^{3/p'-2}\|\nabla M\|_p + \log(\gamma^{-1})\|M\|_{\infty} +\|M\|_2
  +  \|G_2(t)\|_\infty +\|G_2(t)\|_2 \right)
\een
with $0<\gamma \leq 1, 3<p < \infty$. So again each term can be made as small as wished and the proof is complete.
\hfill $\Box$

\section{Proof of the theorem\label{theorem}} 

\medskip
We start by choosing a constant $T_0>0$ such that for $t\geq T_0$ it holds
\be \label{betterdecay}
C_2t^{-9/5} \leq \alpha(1+t)^{-3/2} \quad \text{ and } \quad 
C_2t^{-8/3} \leq \alpha(1+t)^{-5/2},
\ee
where $\alpha$ and $C_2$ are the constants given by Proposition \ref{prop2}. 
Proposition \ref{prop3} says that there exists $\delta >0$ such that a solution
of the Vlasow-Darwin system with initial $f_0 \in \mathcal{D}$ and $\|f_0\|_\infty < \delta$
satisfies
\begin{gather*}
 \|E_L(t)\|_\infty + \|E_T(t)\|_\infty + \|B(t)\|_\infty +
\| \nabla E_L(t)\|_\infty + \| \nabla E_T(t)\|_\infty + \| \nabla B(t)\|_\infty  \\
 < (1+T_0)^{-5/2},
\end{gather*}
for $t$ belonging to $[0,T_0+1]$.  Moreover, it may be assumed that the maximal interval of existence $I = [0,T_{max}[$
is strictly larger than $[0,T_0+1]$, i.e. $T_0+1 < T_{max}$.

If $f$ is a solution as above, then by continuity $f$ satisfies a free streaming condition with parameter $\alpha$
on an interval $[0,T^*]$ with $T_0 < T^* \leq T_{max}$ and $T^*$ may be chosen maximal with these properties. Because of equation (\ref{betterdecay}) we may now conclude with Proposition
\ref{prop2} that
\beas
\|E_L(t)\|_\infty + \|E_T(t)\|_\infty + \|B(t)\|_\infty \leq \alpha (1+t)^{-3/2}, \\
\| \nabla E_L(t)\|_\infty + \| \nabla E_T(t)\|_\infty + \| \nabla B(t)\|_\infty \leq \alpha (1+t)^{-5/2}
\eeas
for all $t \in I$. But this implies $T_{max}= \infty$ and the solution is global.
\hfill $\Box$

%

\section{Spherically symmetric initial data \label{symmetry}}

\medskip
In case the initial datum $f^\circ$ is spherically symmetric, which in the present situation by definition means 
\ben
f^\circ(Qx,Qp)=f^\circ(x,p) \quad \forall x,p \in \R^3, Q \in O(3),
\een
the Vlasov-Darwin system reduces to the relativistic Vlasov-Poisson system, as is shown in the following.
First we show that spherical symmetry is preserved.

\medskip
\begin{lem} 
  Let $f\colon [0,T[\times \R^6 \rightarrow \R$ be a classical solution of the Vlasov-Darwin system and let $f(0)$ be
spherically symmetric. Then $f(t)$ is spherically symmetric for all $0 \leq t<T$.
\end{lem}

\medskip
Actually we will see that any invariance of the initial datum with respect to an orthogonal transformation
is preserved for all times, which implies at the same time that e.g. cylindrical symmetry or reflectional
symmetries are preserved as well.

\medskip
{\em Proof of the Lemma.}

Let $Q \in O(3)$ be given and set $\tilde f(t,x,p) := f(t,Qx,Qp)$. It suffices to show that $\tilde f$ solves the
Vlasov-Darwin system. One finds
\beas
\tilde \rho (t,x) &:=& \int \tilde f (t,x,p)dp = \rho(t,Qx), \\ 
\tilde j(t,x) &:=& \int \tilde f (t,x,p)v(p)dp = Q^{-1}j(t,Qx).
\eeas
For the potentials $\tilde \Phi$ and $\tilde A$ one therefore has
\beas
\tilde \Phi(t,x) &=& \Phi(t,Qx), \\
\tilde A(t,x)&=& Q^{-1}A(t,Qx),
\eeas
as can be seen easily from the Fourier representation of the projection
operator $\mathbb{P}$, compare (\ref{fourier_P}). This implies
\beas
\tilde E_L(t,x) &:= &\nabla \tilde \Phi(t,x) = Q^{-1}E_L(t,Qx), \\
\tilde E_T(t,x) &:= &-\partial_t \tilde A(t,x) = Q^{-1}E_T(t,Qx). \\
\eeas
We set  $\tilde B := \nabla \times \tilde A$. Then by Lemma 2.3 in \cite{pal06} and since
\ben 
\partial_t \tilde \rho(t,x) + \nabla \cdot \tilde j(t,x)=\partial_t \rho(t,Qx) + \nabla \cdot j(t,Qx)=0
\een
the quantities $(E_L,E_T,B)$ solve the field equation part of the Vlasov-Darwin system.

We have to show that the transport equation (\ref{vlasov1}) holds. Consider the term
$p \times \tilde B(t,x) = p \times (\nabla \times \tilde A (t,x))$. By well known
vector identities we can write
\beas
p \times \tilde B(t,x) &=&  \nabla (\tilde A(t,x)p) - (p\cdot \nabla)\tilde A(t,x) ) \\
&=& \left[\left(D\tilde A (t,x)\right)^t - D\tilde A (t,x) \right] p.
\eeas
Here $D$ denotes the total derivative w.r.t. $x$. Now $D \tilde A(t,x)=Q^{-1}DA(t,Qx) Q$
and therefore
\beas
Q(p \times \tilde B(t,x)) &=& \left[ \left(DA(t,Qx) \right)^t - DA(t,Qx)\right] Qp \\
&=& Qp \times B(t,Qx).
\eeas
The last equality holds because the forgoing applies equally well to $A$ as to $\tilde A$.
This finally leads to (\ref{vlasov1}). \hfill $\Box$

\medskip
So we have seen that spherical symmetry is preserved for all times. This includes
that for $t \in [0,T[, Q \in O(3)$ the following identities hold
\beas
\rho(t,Qx) &=& \rho(t,x), \\
\Phi(t,Qx)&=&\Phi(t,x), \\
j(t,Qx) &=& Qj(t,x), \\
A(t,Qx)&=& QA(t,x).
\eeas

\medskip
\begin{lem} The vector field $j$ is radial.
\end {lem}

\medskip
{\em Proof.} In the following the dependence of $j$ on $t$ is suppressed. Let $x \in \R^3\setminus \{0\}$ be given and choose a positive orthonormal basis $(b_1,b_2,b_3)$
with $b_1 = \frac{x}{|x|}$. Let $Q_1, Q_2$ be orthogonal transformations with matrices
\ben
M_1 = \begin {pmatrix}
       -1 & 0 & 0 \\ 0 & -1 & 0 \\0 & 0 & 1
      \end {pmatrix},
\quad 
M_2 = \begin {pmatrix}
       -1 & 0 & 0 \\ 0 & 1 & 0 \\0 & 0 & -1
      \end {pmatrix},
\een
 w.r.t. to the basis chosen. Let $j(x)=\sum_j \alpha_jb_j$. Then
\ben
Q_1j(x)=-\alpha_1 b_1 - \alpha_2 b_2 +\alpha_3b_3, \qquad Q_2j(x)=-\alpha_1b_1 + \alpha_2b_2 - \alpha_3 b_3.
\een
But since $Q_1j(x) =j(Q_1x) = j(-x) = j(Q_2x)=Q_2j(x)$ it follows that $\alpha_2=\alpha_3 = 0$.
\hfill $\Box$

\medskip
\begin{lem} \label{rad_lem}It holds $\mathbb{P}j \equiv 0$.
\end {lem}

\medskip
{\em Proof.} First note, that $\nabla \cdot j(Qx)= \nabla \cdot  j(x)$ for all $Q \in O(3)$. Recall
the definition of $\mathbb{P}$:
\ben
\mathbb{P}j(x) = j(x) + \nabla \Psi(x)
\een
where 
\ben \Psi(x) = \frac{1}{4\pi}\int \frac{\nabla \cdot j(y)}{|x-y|}dy.
\een
But since the source term $\nabla \cdot j$ has rotational symmetry, the forgoing simplifies to
\ben \nabla \Psi(x) = -\int_0^rs^2(\nabla \cdot j)(s) ds \frac{x}{r^3} , \quad r = |x|.
\een
The integral in the last expression can be transformed to
\beas
\int_0^rs^2(\nabla \cdot j)(s) ds & = & \frac{1}{4\pi}\int_{B_r}\nabla \cdot j dV \\
&=&\frac{1}{4\pi}\int_{\partial B_r}jn dS  \\
 &=& j(x)n r^2.
\eeas
So
\ben 
\mathbb{P}j(x) = j(x) + \nabla \Psi(x) = j(x) - j(x)n\frac{x}{r} = 0.
\een 
\hfill $\Box$

\medskip
Lemma \ref{rad_lem} implies that also $A(t)= 0$ for $t \in [0,T[$. This immediately leads to
$E_T = B = 0$, so that the following proposition is proved.

\medskip
\begin{prop}
In case of a spherically symmetric initial datum $f^\circ$ the Vlasov-Darwin system
reduces to the (spherically symmetric) relativistic Vlasov-Poisson system (with repelling forces). 
Hence in this case the solution is global.
\end{prop}

\medskip
The proof of the last statement is given in \cite{gs85}.

\section*{Appendix}

We start with some remarks about the projection operator $\mathbb{P}\colon L^2(\R^3)\rightarrow L^2(\R^3)$, 
which is defined as follows: For $F \in C^1_c(R^3;R^3)$ one sets
\beas
(\mathbb{P}F)(x) &=& F(x)+\nabla \Phi(x),  \qquad \text{ where}\\
\Phi(x) &:=& \frac{1}{4\pi}\int\frac{(\nabla \cdot F)(y)}{|x-y|}dy.
\eeas
Since $-\Delta \Phi = \nabla \cdot F$ we cleary obtain $\nabla \cdot \mathbb{P}F = 0$. Applying the Fourier transform
to these relations it follows
\beas
\hat{\mathbb{P}F}(\xi) & = & \hat F(\xi) + i\hat\Phi(\xi)\xi, \\
\hat\Phi(\xi) &=& \frac{i}{|\xi|^2}\xi \cdot \hat F(\xi),
\eeas
hence
\be \label{frepP}
\hat{\mathbb{P}F}(\xi) = \left( I - \frac{\xi \otimes \xi}{|\xi|^2}\right)\hat F(\xi).
\ee
Therefore $|\hat{\mathbb{P}F}(\xi)|\leq C |\hat F(\xi)|$, so that
by the Plancherel-Theorem $\mathbb{P}$ extends to a continous operator on $L^2(\R^3)$ characterized by (\ref{frepP}).

We conclude with some remarks about the pseudo differential operators used in Section \ref{fields}. Such an operator is of
the form
\be \label{pdo}
Au(x) = \frac{1}{(2\pi)^n}\int A_0(\xi)\hat u(\xi)e^{ix\cdot \xi}d\xi,
\ee
where $n$ is the dimension of the underlying space $\R^n$, the function $A_0$ is called the symbol of the operator and is chosen from a suitable set of functions, and $\hat u$ is the
Fourier transform of $u$. We can restrict ourselves to the case that $u$ belongs to the Schwarz space $\mathcal{S}(\R^n)$ of 
rapidly decreasing functions.
It is shown e.g. in \cite{eskin80} that an operator of the form (\ref{pdo}) with a symbol $A_0$ homogeneous of degree $\alpha > -n$, i.e.
$A_0(t\xi) = t^\alpha A_0(\xi)$ for $t>0,\xi \in \R^n$, has a representation as an integral operator of the form
\[Au(x)= \int a_0(x-y)u(y)dy, \]
where $a_0$ is a function homogeneous of degree $-\alpha - n$. For such a (smooth) function one clearly has
\[
 |a_0(y)| \leq C|y|^{-\alpha-n}
\]
and this is all one needs to know for the estimates presented here and in \cite{ks02}.

\medskip

{\bf Acknowledgement.} I would like to thank Prof. G. Rein for bringing me in contact with this interesting topic and
for many valuable discussions.
\medskip

\bibliographystyle{plain}
\bibliography{literatur}

\begin{thebibliography}{10}

\bibitem{bd85}
C.~Bardos and P.~Degond.
\newblock Global existence for the {V}lasov-{P}oisson system in 3 space
  variables with small initial data.
\newblock {\em Ann. Inst. H. Poincare Anal. Non Lineaire}, 2:101--118, 1985.

\bibitem{ben03}
S.~Benachour, F.~Filbet, P.~Lauren\c{c}ot, and E.~Sonnendr\"ucker.
\newblock {G}lobal existence for the {V}lasov-{D}arwin system in $\mathbb{R}^3$
  for small initial data.
\newblock {\em Math. Meth. Appl. Sci.}, 26:297--319, 2003.

\bibitem{bms07}
N.~Besse, N.~Mauser, and E.~Sonnendr\"ucker.
\newblock {N}umerical approximation of self consistent {V}lasov models for
  low-frequency electromagnetic phenomena.
\newblock {\em Int. J. Appl. Math. Comput. Sci.}, 17(3).

\bibitem{eskin80}
G.I. Eskin.
\newblock {\em Boundary Value Problems for Elliptic Pseudodifferential
  Equations}.
\newblock AMS, Providence, 1980.

\bibitem{gs85}
R.~T. Glassey and W.A. Strauss.
\newblock On symmetric solutions of the relativistic {V}lasov-{P}oisson system.
\newblock {\em Comm. Math. Phys.}, 101:459--473, 1985.

\bibitem{gs87}
R.~T. Glassey and W.A. Strauss.
\newblock Absence of shocks in an initially dilute collisionless plasma.
\newblock {\em Comm. Math. Phys.}, 113:191--208, 1987.

\bibitem{ks02}
S.~Klainerman and G.~Staffilani.
\newblock A new approach to study the {V}lasov-{M}axwell system.
\newblock {\em Commun. Pure Appl. Anal.}, 1(1):103--125, 2002.

\bibitem{pal06}
C.~Pallard.
\newblock {T}he initial value problem for the relativistic {V}lasov-{D}arwin
  system.
\newblock {\em Int. Mat. Res. Not.}, 2006.

\bibitem{r07}
G.~Rein.
\newblock Collisionless kinetic equations from astrophysics---the
  {V}lasov-{P}oisson system.
\newblock {\em Handbook of Differential Equations, Evolutionary Equations. Vol.
  3.}, 2007.

\bibitem{rr92}
G.~Rein and A.D. Rendall.
\newblock Global existence for solutions of the spherically symmetric
  {V}lasov-{E}instein system with small initial data.
\newblock {\em Comm. Math. Phys.}, 150:561--583, 1992.

\bibitem{sg06}
H.~Schmitz and R.~Grauer.
\newblock {D}arwin-{V}lasov simulations of magnetized plasmas.
\newblock {\em arXiv:physics/0601220v1}, 2006.

\end{thebibliography}

\end{document}